\documentclass[twocolumn, amsmath,superscriptaddress, amsfonts,prb]{revtex4}
\usepackage{graphicx}
\usepackage{url}
\begin{document}

\title{Hot spot runaway in thin film photovoltaics and related structures}
\author{V. G. Karpov}\email{vkarpov@physics.utoledo.edu}\affiliation{Department of Physics and Astronomy, University of Toledo, Toledo, OH 43606, USA}
\author{A. Vasko}\affiliation{Department of Physics and Astronomy, University of Toledo, Toledo, OH 43606, USA}
\author{A. Vijh}\affiliation{Xunlight Corporation, 3145 Nebraska Avenue, Toledo, OH 43607, USA}
\begin{abstract}

We show that thin film diode structures, such as photovoltaics and light emitting arrays, can undergo zero threshold localized thermal runaway  leading to thermal and electrical nonuniformities spontaneously emerging in originally uniform systems. The linear stability analysis is developed for a system of thermally and electrically coupled two discrete diodes, and for a distributed system. These results are verified with numerical modeling that is not limited to small fluctuations. The discovered instability negatively affects the device performance and reliability. It follows that these problems can be mitigated by properly designing the device geometry and thermal insulation.

\end{abstract}

\date{\today}

\maketitle

Many thin film devices, such as photovoltaics or light emitting arrays, are thermally insulated and have low lateral thermal conduction. As a result, temperature fluctuations in them can be relatively long lived. With the current-voltage characteristics and power generation of a diode having exponential dependence on temperature, this can lead to the thermal runaway instability.  The mechanism of it is that local temperature fluctuation creates a spot of increased transversal conduction, through the diode, that acts as a shunt robbing currents from the surrounding area; this further increases local heat power generation and temperature.

The phenomenon of runaway instability has long been known in electrical engineering (current hogging), chemistry (temperature accelerated exothermic reactions), astrophysics (nova explosions due to runaway nuclear fusion), and other fields where increase in temperature causes positive feedback. Understanding of runaway instability is most advanced for thermal explosions. \cite{frank1969,zeldovich1985,kotoyori2005} In particular, it was realized that the instability starts with a hot spot representing a finite amplitude local temperature fluctuation \cite{merzhanov1971,thomas1973} resembling nucleation processes in phase transitions of the first kind. This analogy was elegantly explored in Ref. \onlinecite{subashiev1987} and more recently discussed for thin film structures. \cite{karpov2012}

Here we consider the runaway thermal instability in thin film diode systems of practical interest. We show that they can be unstable with respect to infinitesimally small fluctuations (as opposed to the above mentioned examples \cite{frank1969,zeldovich1985,kotoyori2005,merzhanov1971,thomas1973,subashiev1987,karpov2012}). Remarkably, the latter property can be established already for a basic model of two identical coupled discrete diodes in parallel, with which we start our analysis. A distributed diode model discussed next leads to the same conclusion.
\begin{figure}[t!]
\includegraphics[width=0.35\textwidth]{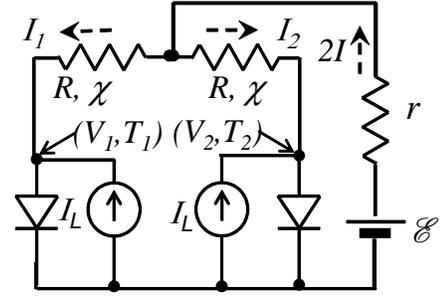}
\caption{The two-diode model.  \label{Fig:2diode}}
\end{figure}

{\it The two-diode model: linear stability analysis.} Fig. \ref{Fig:2diode} presents a model of two identical diodes of heat capacitance $C$ each, coupled electrically through a resistor $2R$, and thermally through a connector of thermal conduction $\chi$. The load resistor $r$ and the power source ${\cal E}$ form an external part of the circuit governing the total voltage and current through the diodes. We assume the standard activated saturation currents
\begin{equation}\label{eq:satcur} I_{01(2)}=I_{00}\exp(-E/kT_{1(2)}),\end{equation}
and the diode current voltage characteristics,
\begin{equation}\label{eq:diodeIV}
I_{1(2)} = I_{01(2)}[\exp(qV_{1(2)}/kT_{1(2)})-1]-I_L,\end{equation}
where $V_{1(2)}$, $I_{1(2)}$ and $T_{1(2)}$ are respectively the voltage, current and temperature of the diode 1(2), $q$ is the elemental charge, and $I_L$ is the photocurrent.
Considering the current through resistors $R$ and $r$ one gets in addition,
\begin{equation}\label{eq:resR}V_1+I_1R=V_2+I_2R={\cal E}-2Ir. \end{equation}

The heat transfer is described by the equation
\begin{equation}\label{eq:heat}C\frac{dT_{1(2)}}{dt}=H+I_{1(2)}V_{1(2)}-\chi (T_{1(2)}-T_{2(1)})-\alpha (T_{1(2)}-T_0)
\end{equation}
where $H$ represents uniform heat generation due to direct light absorption,  $T_0$ is the ambient temperature, and $\alpha$ is the coefficient in the Newton cooling law. Here the Joule heating ($IV$ term) for the diode can be substantiated by noting that the energy liberated is the difference between that of absorbed photon and the final energy of the electron and hole at their quasi-Fermi levels, $F_e$ and $F_h$. The latter contains the difference $F_e-F_h=qV$ where $V$ is the voltage across the diode.

Let us perform the linear stability analysis of the above standard equations. Because the diodes are identical, their unperturbed currents, voltages, and temperatures are the same,
\begin{equation}\label{eq:sym}
I_1=I_2=I,\quad V_1=V_2=V,\quad  T_1=T_2=T.
\end{equation}
In the next approximation, consider small variations,
\begin{equation}\label{eq:delta}
T_{1(2)}=T+\delta T_{1(2)},  V_{1(2)}=V+\delta V_{1(2)}, I_{1(2)}=I+ \delta I_{1,2}.\nonumber
\end{equation}
They are coupled through  Eqs. (\ref{eq:diodeIV}) and (\ref{eq:resR}),
\begin{equation}\label{eq:interm}
\frac{\delta V_2-\delta V_1}{\delta I_2-\delta I_1}=-R,\quad \delta V_{1,2}=\frac{k\delta T_{1,2}}{q}{\cal L}+\frac{kT_{1,2}\delta I_{1,2}}{I+I_L}
\end{equation}
where
\begin{equation}\label{eq:calL}
{\cal L}\equiv \ln\left((I+I_L)/I_{00}\right).\nonumber
\end{equation}

Subtracting equations (\ref{eq:heat}) for the two diodes from each other and using Eqs. (\ref{eq:interm}), the temperature asymmetry is described by,
\begin{equation}\label{final1}
d\delta T/dt=-\Gamma \delta T,\quad \delta T\equiv\delta T_1-\delta T_2
\end{equation}
where
\begin{equation}\label{eq:final2}
\Gamma =\frac{1}{C}\left\{2\chi +\alpha +\frac{k(I+I_L)(V-IR){\cal L}}{q[R(I+I_L)+(kT/q)]}\right\}.
\end{equation}

It follows that fluctuations in the temperature (and thus electric) asymmetry can decay ($\Gamma >0$) or spontaneously grow ($\Gamma <0$) depending on the system parameters. Note in this connection that typically ${\cal L}$ is negative and large in absolute value (see below); hence, the third term on the right hand side in Eq. (\ref{eq:final2}) can be negative and dominating, in which case the fluctuations spontaneously grow.

The interplay between the stable and unstable regimes depends on the parameters in Eq. (\ref{eq:final2}) where values of $V$ and $I$ are governed by the entire circuit. It follows from Fig. \ref{Fig:2diode} that
\begin{equation} \label{eq:outer} V+IR+2Ir-{\cal E}=0, \quad I = I_{0}[\exp(qV/kT)-1]-I_L.\end{equation}
The analysis of these equations is straightforward leading to a variety of cases; we will briefly mention few of them. \\(a) Open circuit regime: $r=\infty$, $I=0$, $V=(kT/q)\ln(I_L/I_0)$. Depending on realistic values of other parameters (see below), the system can be either unstable or stable. \\(b) Short circuit regime: $V=0$, $I=-I_L$. The third term on the right hand side of Eq. (\ref{eq:final2}) disappears, and the system is stable. \\(c) Strong forward bias: $I\gg I_L$, $RI\gg kT/q$, $V-IR\approx 2I_0r\exp(qV/kT)+{\cal E}$. The third term in Eq. (\ref{eq:final2}) becomes negative and exponentially large in absolute value leading to instability. \\(d) Strong reverse bias: $V-IR\approx V<0$, $I+I_L\approx -I_0$, $I_0R\ll kT/q$. The third term in Eq. (\ref{eq:final2}) is positive (estimated as $-I_0V{\cal L}/T$) and the system stable.

It is worth noting that the instability (current hogging) in parallel identical diodes has long been known as a practical issue reflected in  multiple internet blogs, \cite{blog1,blog2,blog3,blog4,blog5} though it has been typically attributed to inevitable variations in device parameters rather than an instability that arises in even perfectly identical devices. The role of the above analysis is that it provides a quantitative model, defines the essential physical quantities, and establishes the instability parameter $\Gamma$.
\bigskip

\begin{figure}[t!]
\includegraphics[width=0.35\textwidth]{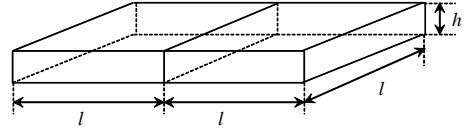}
\caption{A simplified thin film geometry of two neighboring areas in a thin film structure representing the two-diode model.  \label{Fig:geom}}
\end{figure}

{\it The two-diode modeling of thin-film structures: numerical estimates and scaling.}
The two-diode model can be qualitatively related to a diode thin film structure geometry as illustrated in Fig. \ref{Fig:geom}. Using the standard definition of parameters, that geometry yields,
\begin{eqnarray}\label{eq:scaling}
R&=&const,\quad (I,I_L,I_{00})=(j,j_L,j_{00})l^2,\quad V=const,\nonumber \\
\alpha &=&\alpha _0l^2,\quad \chi =\chi _0h,\quad {\cal L}=const.\nonumber
\end{eqnarray}
Here $const$ can include logarithmical dependencies as negligibly weak compared to power dependencies on $l$ and $h$ with other quantities, $j$'s with indices stand for the corresponding current densities (A/cm$^2$).

We use the typical thermal conductivity $\chi _0\sim 10$ W/m-K and the heat transfer coefficient from the Newton's cooling law $\alpha _0\sim 10-10^3$ W/m$^2$-K (for various thermal contacts). Also, we assume the typical $R\sim 1-10$ Ohm, $j_L\sim 20$ mA/cm$^2$, $j\approx -0.9j_L$ (at the maximum power point). The parameter ${\cal L}\sim -100$ is estimated in the order of magnitude from the existing data (see e. g. Refs. \onlinecite{liang2005, liu2012,singh2012}) on the temperature dependence of the open circuit voltage, $V_{oc}=E+(kT/q){\cal L}$.

Consider first the case of $l=1$ cm and $h=10 $ $\mu$m, in the ballpark of practical parameters, for which $\chi \sim 10^{-4}$ W/K, $\alpha\sim 10^{-3}-0.1$ W/K, and the third term in Eq. (\ref{eq:final2}) is of the order of $-2\cdot 10^{-3}$ W/K; hence, the term in parentheses in Eq. (\ref{eq:final2}) can be negative, i. e. fluctuations grow. Large non-ideality factors (effectively increasing $kT/q$), low sheet resistances ($R$), and effective thermal insulation (low $\alpha$) are conducive to such instabilities.

The size for practical, efficient devices is limited by the screening length\cite{Luk,Kar2}
\begin{equation}\label{eq:scr}
L=\sqrt{V_{oc}/|j|R}\sim 1\quad {\rm cm},
\nonumber \end{equation}
allowing current collection without significant resistive loss. Note the characteristic length,
\begin{equation}
l_c=\sqrt{\frac{kT}{q(j+j_L)R}}=L\sqrt{\frac{kT}{qV_{oc}}\frac{|j|}{j+j_L}},
\label{eq:lc}\end{equation}
such that when $l<l_c$ the denominator of the last term in Eq. (\ref{eq:final2}) changes its sign and the instability becomes impossible. Since typically $kT\ll qV_{oc}$, one can expect $l_c<L$, i. e. the range of instabilities exists.

However, the instability may cease to exist even for $l>l_c$ if the cell is small enough. Consider for example $l=1$ mm, all other parameters remaining the same. This decreases the second and third terms on the right-hand-side of Eq. (\ref{eq:final2}) by two orders of magnitude; hence, $\Gamma >0$.

We conclude that, the runaway instability is possible in a finite cell size range (typically mm scale) limited by strong thermal contact from below and by resistance loss from above.

\begin{figure}[t!]
\includegraphics[width=0.42\textwidth]{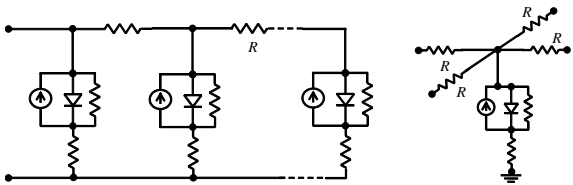}
\caption{Left: A one-dimensional finite element representation of the distributed diode model where each node includes additional shunt and series resistances. Right: a single node structure of a two-dimensional distributed diode model.   \label{Fig:2diodedist}}
\end{figure}
\bigskip

{\it Distributed diode model}
The distributed model (Fig. \ref{Fig:2diodedist}) electric and heat transport equations take the form
\begin{eqnarray}\label{eq:distr}
\nabla ^2V&=& Rj(V,T),\nonumber \\
c\frac{\partial T}{\partial t} &=& H+jV+\chi \nabla ^2T-\alpha (T-T_0).
\end{eqnarray}
The former describes current branching through diode elements between the resistive and ground (ideal) conductor; \cite{Luk,Kar2} the latter has the same meaning as Eq. (\ref{eq:heat}). Here $\nabla ^2$ stands for the two-dimensional Laplacian in the lateral directions, $H$ is the uniform heat per area, $c$ is the specific heat per area, and $\chi _0 \sim h\chi _0$ has the dimensionality of $W/K$. The diode-type current voltage characteristic expresses current density (A/m$^2$), e. g.
\begin{equation}\label{eq:jV}
j=j_{00}\exp(-E/kT)[\exp(qV/kT)-1]-j_L,\nonumber
\end{equation}
or similar, say, including additional shunt and series resistances. For such characteristics, one has
\begin{equation}\label{eq:deriv}
\partial j/\partial V\equiv j_V'>0\quad {\rm and}\quad \partial j/\partial T\equiv j_T'>0.\nonumber
\end{equation}

Along the lines of the standard linear stability analysis, consider small perturbations $$\delta V=v\exp(i{\bf k\cdot x}+i\omega t) \quad {\rm and}\quad \delta T=\theta \exp(i{\bf k\cdot x}+i\omega t)$$ to the average uniform values $V$ and $T$ respectively, and their related current fluctuation,
$$\delta j=j_V'\delta V+j_T'\delta T.$$ Eqs. (\ref{eq:distr}) then take the form
\begin{eqnarray}\label{eq:ampl}
(k^2+Rj_V')v+Rj_T'\theta &=&0, \nonumber \\
-(j+Vj_V')v+(ic\omega +\chi k^2 +\alpha -Vj_T')\theta &=&0.\nonumber
\end{eqnarray}
Setting the determinant of the latter system equal to zero yields the dispersion equation
\begin{equation}\label{eq:disp}
i\omega =-\frac{\chi k^4-Ak^2+B}{c(k^2+Rj_V')}
\end{equation}
allowing instability ($i\omega >0$) in the domain $k_-<k<k_+$ where
\begin{equation}\label{krange}
k_{\pm}=\left(\frac{A\pm\sqrt{A^2- 4\chi B}}{2\chi}\right)^{1/2},
\end{equation}
\begin{equation}\label{eq:AB}
A=Vj_T'-\chi Rj_V'-\alpha, \quad B = R(\alpha j_V'+jj_T')  .\nonumber
\end{equation}

The domain of instability exists when $A>0$, a condition that imposes restrictions on the values of heat transfer parameter $\alpha$ and $\chi$, qualitatively similar to the result in Eq. (\ref{eq:final2}) for the model of two discrete diodes. The case of low heat transfer (neglecting $\alpha$ and $\chi$ where possible), can be easily analyzed more quantitatively,
\begin{eqnarray}\label{eq:param}
k_+&\approx &\sqrt{\frac{A}{\chi}}\sim \sqrt{\frac{Vj_T'}{\chi}}\sim \sqrt{\frac{V(j+j_L)E}{kT^2\chi}}\sim 1\quad {\rm mm}^{-1},\nonumber \\
k_-&\approx &\sqrt{\frac{B}{A}}\sim \sqrt{\frac{Rj_T'j}{Vj_T'}}= \sqrt{\frac{Rj}{V}}\sim 1\quad {\rm cm}^{-1},\nonumber
\end{eqnarray}
where we have used the same parameter values as in the above. These estimates are in fair agreement with the earlier results obtained by the discrete two diode model; in particular, $k_-\sim 1/L$. Overall, the originally uniform device is predicted to spontaneously break into a system of domains of linear dimension $k^{-1}$, each containing a hot spot caused by the runaway instability. We note that the distributed model should be compared with the open circuit regime of the above two-diode model, since no current collection is specified in the former.

\begin{figure}[t!]
\includegraphics[width=0.38\textwidth]{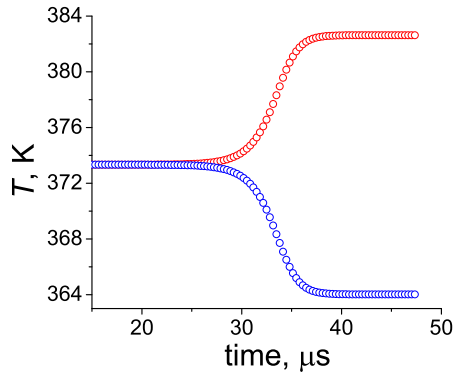}
\caption{Numerical modeling of a runaway instability in a system of two diodes using $T_0=290$ K, $\chi = 10^{-5}$ W/K, $\alpha =1.2\cdot 10^{-3}$ W/K, and $R=0.1$ $\Omega$. These parameters are similar to the estimates given above, and lead to an equilibrium temperature for the two diodes (at zero time) of 370 K, similar to the equilibrium temperature of real cells under light soak. \cite{cdtetemp} Their temperatures become different over time, following the two branches that correspond to the heating and cooling diodes, and eventually saturate. \label{Fig:modeling}}
\end{figure}
\bigskip

{\it Numerical modeling.} Our numerical modeling dealt with a finite number of photo-diodes including additional series and parallel resistances as illustrated in Fig. \ref{Fig:2diodedist}, where the nearest neighbors were connected through resistive electrodes and finite thermal conductors, assuming as well the Newton's cooling law. In addition to verifying the above analytical results, that modeling was able to track the thermal runaway in time following its saturation, as illustrated in Fig. \ref{Fig:modeling}. The saturation temperature and time are governed mostly by the thermal exchange coefficient $\alpha$ and the ambient temperature $T_0$. In the range of small fluctuations, our modeling results are consistent with the preceding analytical consideration.

We would like to mention two implications of the runaway related nonuniformities. (1) They make it impossible for the entire device area to operate under the same optimum power condition; hence, there is a decrease in device performance. (2) Nonuniform material degradation accelerates at hot spots, such that an initial hot spot may then degrade in a runaway mode under more and more stress as it becomes progressively more shunting. The final result of such degradation will be roughly one shunt per the area of linear dimension $L$. It is remarkable that the above mentioned problem with performance and reliability related to local spot runaway instability can be fixed by properly scaling the device thickness, substrate material, and thermal insulation.

A comment is in order regarding the very fact that we have here discovered zero threshold instability, which may appear inconsistent with the conclusion of previous work \cite{frank1969,zeldovich1985,kotoyori2005,merzhanov1971,thomas1973,subashiev1987,karpov2012} that runaway can start with finite amplitude temperature fluctuations. This seeming inconsistency can be attributed to a rather specific combination of the diode current voltage characteristic (I/V) and its related heat generation written in the quasi-Joule form, $H+IV$. As a result, the heat generation is not simply thermally activated (which was assumed in all previous work), but the activation energy becomes a nontrivial function of temperature. However, the present analysis by no means rules out finite threshold fluctuations in the systems under consideration; they will be analyzed elsewhere.

In conclusion, we have shown that, under certain conditions, thin-film diode-like structures possess hot spot runaway instabilities of zero threshold. This was proven by the linear stability analyses of two discrete diode model and distributed diode model, and verified by direct numerical simulations. The instabilities are favored by low heat transfer parameters (thermal conductivity, thickness, and Newton's cooling law coefficient) as well as low sheet resistances and significant currents and voltages. These factors can be tweaked to mitigate the detrimental effects of such instabilities which are loss in performance and nonuniform device degradation.

This work was performed under the auspice of the NSF award No.  1066749.
Discussions with A. V. Subashiev and D. Shvydka are greatly appreciated.

\end{document}